\documentclass[aip,pof,12pt]{revtex4-1}
\usepackage{graphicx}
\usepackage{amssymb,amsmath}
\usepackage{natbib}
\usepackage{subfigure}
\usepackage{epstopdf}
\usepackage{color}

\newcommand{\nom}{Nu_\omega}
\newcommand{\etal}{\emph{et al.}}

\begin{document}

\title{Effects of the computational domain size on DNS of Taylor-Couette turbulence}

\author{Rodolfo Ostilla-M\'{o}nico$^1$, Roberto Verzicco$^{1,2}$ and Detlef Lohse}
\affiliation{
$^{1}$ Physics of Fluids Group, Department of Science and Technology, Mesa+ Institute,  and J.\ M.\ Burgers Centre for
Fluid Dynamics,\\ University of Twente, 7500 AE Enschede, The Netherlands \\
$^2$Dipartimento di Ingegneria Industriale, University of Rome ``Tor Vergata'', Via del Politecnico 1, Roma 00133, Italy }
\date{\today}

\begin{abstract}
In search for the cheapest but still reliable numerical simulation, a systematic study on the effect of the computational domain (``box'') size on direct numerical simulations of Taylor-Couette flow was performed. Four boxes, with varying azimuthal and axial extents were used. The radius ratio between the inner cylinder and the outer cylinder was fixed to $\eta=r_i/r_o=0.909$, and the outer was kept stationary, while the inner rotated at a Reynolds number $Re_i=10^5$. Profiles of mean and fluctuation velocities are compared, as well as autocorrelations and velocity spectra. The smallest box is found to accurately reproduce the torque and mean azimuthal velocity profiles of larger boxes, while having smaller values of the fluctuations than the larger boxes. The axial extent of the box directly reflects on the Taylor-rolls and plays a crucial role on the correlations and spectra. The azimuthal extent is also found to play a significant role, as larger boxes allow for azimuthal wave-like patterns in the Taylor rolls to develop, which affects the statistics in the bulk region. For all boxes studied, the spectra does not reach a box independent maximum.
\end{abstract}

\maketitle

\section{Introduction}

Taylor-Couette (TC), the flow between two coaxial and independently rotating cylinders, is commonly used as a basic model and paradigmatic system for shear flows in very diverse topics, for example probing the stability of astrophysical flows \cite{ji13} or more direct applications such as bubbly drag reduction \cite{ber05,gil13}.  TC is very accessible experimentally, as it is a closed system, it has a high number of symmetries and a simple geometry. Experiments of TC have been conducted up to $Re\sim\mathcal{O}(10^6)$ \cite{lat92,lat92a,gil11a,gil11}. These large $Re$ allow for the study of the ``ultimate'' regime, in which the flow is fully turbulent both in the boundary layers and in the bulk. It is expected that the scaling laws which hold in this regime can be extrapolated to arbitrarily large Reynolds numbers, such as those present in geo- and astro-physics \cite{kra62,gro11}. 

The flow is sheared by the angular velocity difference between the two cylinders. The driving of the cylinders can be expressed non-dimensionally with two Reynolds numbers: $Re_i=r_id\omega_i/\nu$ for the inner cylinder and $Re_o=r_od\omega_i/\nu$ for the outer cylinder, where $r_i$ and $r_o$ are the inner and outer cylinder radius, $\omega_i$ and $\omega_o$ the inner and outer cylinder angular velocity, $d$ is the gap width, $d=r_o-r_i$ and $\nu$ the kinematic viscosity of the fluid. The shear driving of the flow can then be expressed as a shear Reynolds number $Re_s = r_i|\omega_i-\omega_o|d/\nu = |Re_i-\eta Re_o|$, where $\eta$ is the radius ratio $\eta=r_i/r_o$.

Direct numerical simulations (DNS) of TC have received increasing level of sophistication in the last years, in an attempt to reach the ultimate regime numerically. The first attempt at high $Re$ was done using ``large'' computational boxes by Dong \cite{don07,don08}. Dong used a periodic aspect ratio of $\Gamma=2\pi$, where $\Gamma=L/d$, with $L$ the axial periodicity length, for $\eta=0.5$, reaching $Re_s=8000$. Ostilla-M\'onico \etal\cite{ost13,ost14c} also achieved $Re_s\approx 8000$ for $\eta=0.714$ using large boxes with $\Gamma=2\pi$. A breakthrough was achieved by Brauckmann \& Eckhardt \cite{bra13,bra13b}, who showed that simulation boxes could be heavily reduced in two ways, while still obtaining accurate data for the torque. For simulation boxes with $\Gamma=2\pi$, three Taylor roll pairs fit in the system. However, only one pair of rolls was sufficient to calculate the torque, so $\Gamma$ could be reduced to $\Gamma=2$ ($\approx2\pi/3$). Secondly, simulating the full azimuthal extent of the cylinder is also not necessary to obtain an accurate result for the torque. A cylindrical wedge, with a rotational symmetry can be imposed, and, for $\eta=0.714$, only a \emph{ninth}\cite{bra13} of the cylinder was necessary. The use of these ``small'' boxes reduces the computational requirements by a factor $\sim 30$, or more, and made later DNS deep inside the ultimate regime by Ostilla-M\'{o}nico \etal \cite{ost14,ost14e} possible, who achieved $Re_s\sim\mathcal{O}(10^5)$.

The torque at the inner and outer cylinders are a first order, one-point statistic. The finiteness of the computational domain however may play a role for other statistics, both higher order statistics, or two- and many-point statistics. This is the case for example in the channel flow simulations of Lozano-Dur\'{a}n \etal \cite{loz14}, where even if the stream-wise mean velocity profile is well reproduced with small boxes, accurate results for velocity and pressure fluctuations (root mean squared) require larger boxes. 

Before we continue, it is worth noting that there are two main difference between channels (and pipes) and TC. In TC, a natural constraint on the azimuthal (streamwise) extent of the domain always exists, i.e. the full cylinder, while one could think of channels and pipes extending infinitely. Second, the axial (spanwise) periodicity length in TC fixes the size of the Taylor rolls. The effect of the box-size in the axial direction is not purely numerical, as the wavelength of the Taylor vortex is a \emph{physical} parameter. These rolls survive up to very large Reynolds numbers, having been observed experimentally up to $Re_s\sim\mathcal{O}(10^6)$ by Huisman \etal\cite{hui14}, and, in the corresponding parameter regimes all DNS simulations up to now. Taylor rolls have no direct analog in pipes and channel flow between two parallel plates, and can still play a large role in the DNS for various physical quantiites, even though the torque has been shown to become independent of $\Gamma$ in the range $2\leq\Gamma\leq4$ at about $Re_s\sim 3\cdot 10^4$~~\cite{ost14e}. 

In this manuscript we attempt to answer the question: how does the size of the computational domain affect other statistics of TC DNSs and in particular higher order moments? To do so, we performed a series of DNSs of TC using a second-order finite difference code, with fractional time-stepping detailed in Verzicco \& Orlandi. \cite{ver96} This code has been used for all previous DNS of TC, and has been extensively validated against experiments. \cite{ost13,ost14,ost14e} 

The radius ratio was fixed to $\eta=0.909$, the inner cylinder was rotated at $Re_i=10^5$, while the outer cylinder was kept stationary, i.e. $Re_o=0$. This resulted in a total shear driving of $Re_s=10^5$. With the chosen parameters, the simulations are in the fully turbulent (ultimate) regime, and still have a strong large-scale axial circulation \cite{ost14e}. The flow is fully Rayleigh unstable, i.e. $d|\omega r^2|/dr < 0$ everywhere.  For these conditions, four simulations were conducted with computational boxes of varying sizes. Details of the geometry and resolutions used are available in Table~\ref{tbl:final}. The adequacy of the mesh can be further checked in the spectra shown in later sections. After a sufficiently long time to let transient behaviour die out, simulations were run for about 30 large eddy turnover times based on $d/(r_i\omega_i)$.

Throughout this manuscript, the following conventions will be used: $\langle \phi \rangle_{x_i}$ denotes the average of a quantity $\phi$ with respect to the independent variable $x_i$. The torque $T$ is non-dimensionalized as a pseudo-Nusselt number \cite{eck07b} $\nom=T/T_{pa}$ where $T_{pa}$ is the torque in the purely azimuthal and laminar state. We also define $\tilde{r}$, the normalized radius as $\tilde{r}=(r-r_i)/d$, the normalized height as $\tilde{z}=z/d$, and the normalized azimuthal distance at the mid-gap as $\tilde{x}=(r_o+r_i)\theta/(2d)$.  

Normalizations with respect to ``wall'' variables are denoted with a plus superscript, i.e. $\phi^+$. Wall variables are first averaged azimuthally, axially and temporally. Then, the frictional velocity at the corresponding cylinder $u_\tau = \sqrt{\tau_w/\rho}$ is computed, and used as velocity scale, where $\tau_w$ is the mean friction at the corresponding cylinder, and $\rho$ is the fluid density. As length scale to non-dimensionalize the viscous length $\delta_\nu = \nu/u_\tau$ is used as usual.  In these wall variables, we denote the distance to the cylinder(s) with $r^+$. For the inner cylinder wall variables, this is defined as $r^+=(r-r_i)/\delta_{\nu,i}$, while for the outer cylinder wall variables $r^+=(r_o-r)/\delta_{\nu,o}$. 

\begin{table}[htp]
  \begin{center} 
  \def~{\hphantom{0}}
  \begin{tabular}{|c|c|c|c|c|c|c|}
  \hline
  Case & $n_{sym}$ & $\Gamma$ & $N_\theta\times N_r \times N_z$ & $\nom$ & $Re_{\tau,i}$ & Line Style \\
  \hline
  $\Gamma 2\text{N}20$ & $20$ & $2.09$ & $1024 \times 1024 \times 2048$ & $69.5\pm0.2$ & $1410$ & Solid light blue \\
  $\Gamma 2\text{N}10$ & $10$ & $2.09$ & $2048 \times 1024 \times 2048$ & $69.4\pm0.4$ & $1410$ & Solid black \\
  $\Gamma 3\text{N}20$ & $20$ & $3.00$ & $1024 \times 1024 \times 3072$ & $69.6\pm0.2$ & $1410$ & Dashed dark green \\
  $\Gamma 4\text{N}10$ & $10$ & $4.00$ & $2048 \times 1024 \times 4096$ & $69.8\pm1.6$ & $1410$ & Dash-dot dark red \\
  \hline
 \end{tabular}
 \caption{Details of the numerical simulations. The first column is the name with which the simulation will be refereed to in the manuscript. The second column shows $n_{sym}$, the order of the rotational symmetry imposed on the system. The third column gives $\Gamma$, the axial periodicity aspect ratio. The fourth column represents the amount of points in the azimuthal, radial and axial directions used for the simulations. The resolutions used correspond to approximately $r_o\Delta\theta^+\approx9$ and $\Delta z^+\approx3$ in inner cylinder wall units. The fifth column shows the non-dimensional torque $\nom$. The sixth column displays $Re_{\tau,i}=u_{\tau,i}d/(2\nu)$, the frictional Reynolds number at the inner cylinder. $Re_{\tau,o}$ can be obtained from $Re_{\tau,o}=\eta Re_{\tau,i}$. The last column indicates the line shapes used for Figs. \ref{fig:q1wall}-\ref{fig:spectraslab9}.}
 \label{tbl:final}
\end{center}
\end{table}

\begin{figure}
  \centering
  \includegraphics[trim=4cm 0cm 3cm 0cm,clip=true,height=6cm,angle=-0]{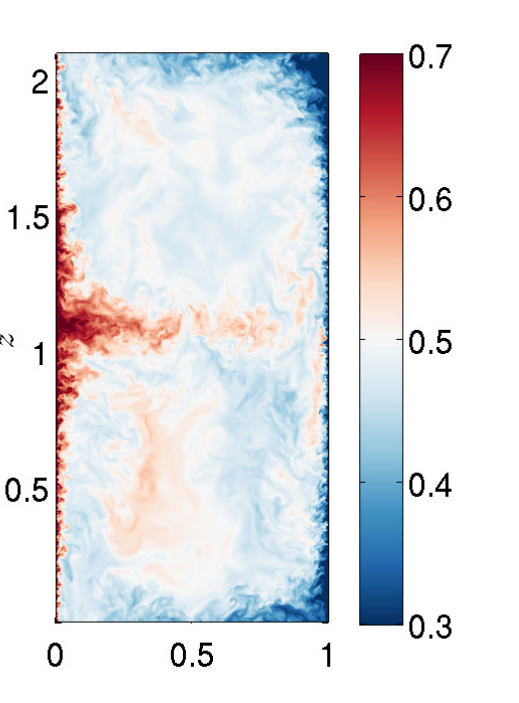}
  \includegraphics[trim=4cm 0cm 3cm 0cm,clip=true,height=6cm,angle=-0]{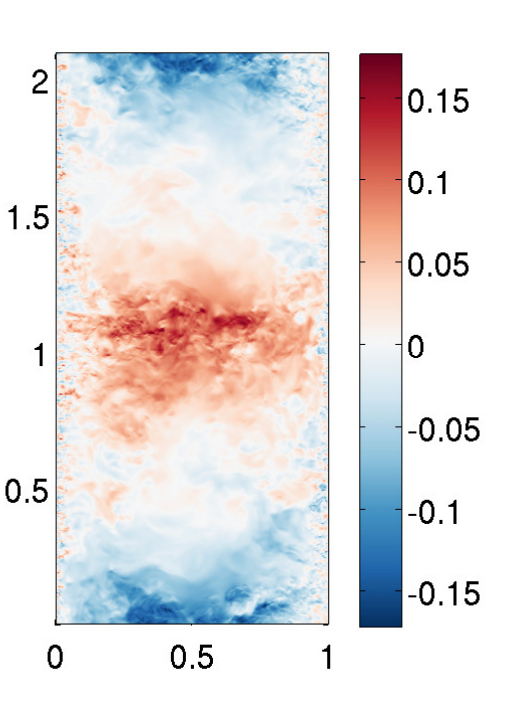}
  \includegraphics[trim=4cm 0cm 3cm 0cm,clip=true,height=6cm,angle=-0]{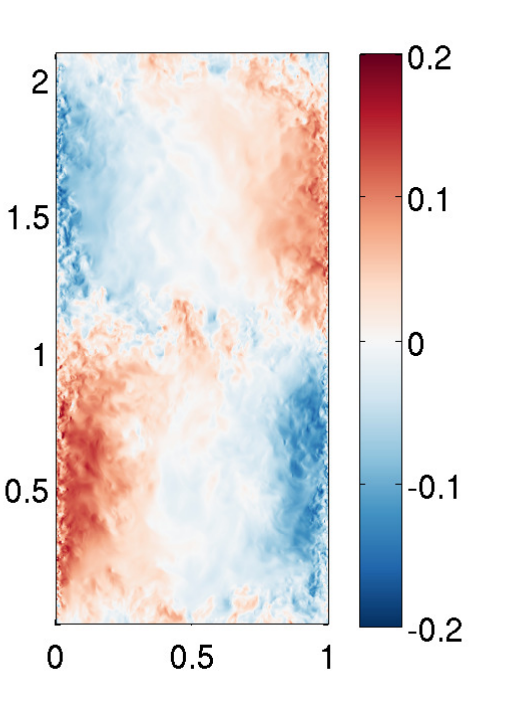}
  \caption{ Pseudocolor of the instantaneous azimuthal (left), radial (middle) and axial (right) velocities for an azimuthal cut of $\Gamma2\text{N}20$. The presence of large scale rolls, which make the flow inhomogeneous, is apparent here.  }
\label{fig:stst4cut}
\end{figure}

Table~\ref{tbl:final} also shows that $\nom$ is the same within the statistical temporal error due to the necessarily limited time averaging for all simulations. This finding, even if expected from previous research, is still remarkable, considering the large-scale flow patterns, i.e. the Taylor rolls, which are still present in the flow. These patterns can be appreciated from Fig.~\ref{fig:stst4cut}. Structures are emitted from the boundary layers. These can be thought of as hairpin vortices, or as plumes when speaking in the language common for thermal convection. Plumes tend to attract each other, and merge together forming regions with large angular velocity transport, i.e. with a strong correlation between $u_\theta$ and $u_r$. These regions have a very large positive or negative angular velocity transport, which can be orders of magnitude larger than the mean \cite{ost14}. ``Neutral''-transport regions, in the core of the Taylor rolls, lie between them, and in these regions $u_r$ and $u_z$ are small. For a more detailed analysis of the dynamics of these regions, see Refs.\cite{ost14},\cite{ost15}.

\begin{figure}
  \centering
  \includegraphics[trim=4cm 0cm 4cm 0cm,clip=true,height=6cm,angle=-0]{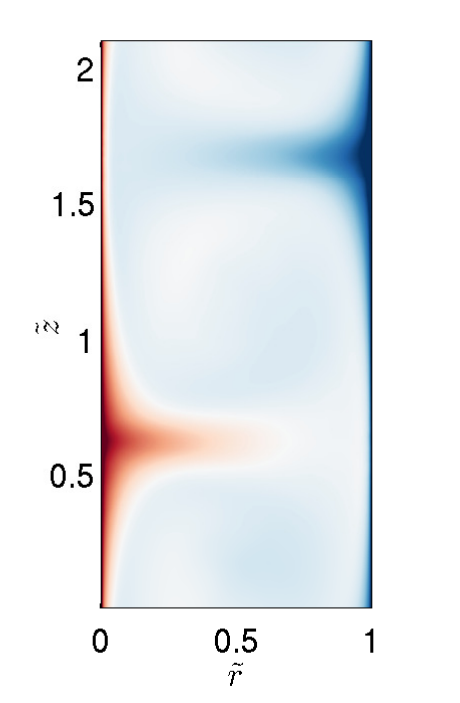}
  \includegraphics[trim=4cm 0cm 4cm 0cm,clip=true,height=6cm,angle=-0]{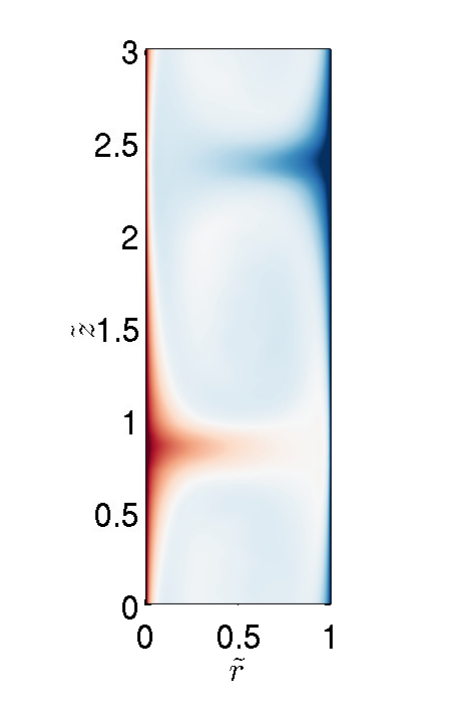}
  \includegraphics[trim=4cm 0cm 4cm 0cm,clip=true,height=6cm,angle=-0]{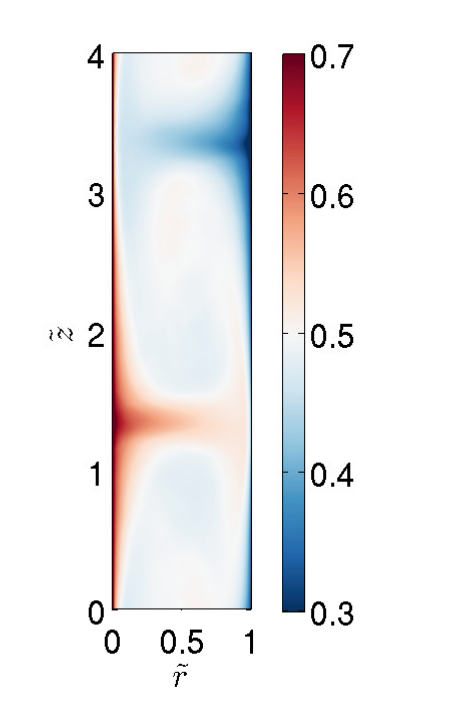}
  \caption{ Azimuthally- and time- averaged azimuthal velocity $\bar{u}_\theta$ for the cases $\Gamma 2\text{N}20$ (left), $\Gamma 3\text{N}20$ (middle) and $\Gamma 4\text{N}10$ (right). As $\Gamma$ increases, the Taylor roll pair also grows, filling the entire box. In these panels (and also in Figs.~\ref{fig:stst4cut} \& \ref{fig:q1insttheta}) there is a slight preference for blue (low velocity) regions in the plots, because the mean azimuthal velocity in the bulk is slightly below $0.5$ due to the inherent asymmetry between both cylinders. }
\label{fig:q1me}
\end{figure}

Taylor-rolls are stationary in time. This is displayed by Fig.~\ref{fig:q1me}, which shows a pseudocolour plot of the azimuthally- and temporally- averaged azimuthal velocity $\langle u_\theta\rangle_{\theta,t}$. For all panels, a single vortex pair is present, which fills up the whole computational domain. This large-scale structure has little to no effect on the total angular velocity transport. The simulations were ran for more than $30$ large eddy turnover times (defined as $\tilde{t}=d/r_i\omega_i$), and this did not significantly modify the position of the roll. Remarkably, for the $\Gamma4\text{N}10$ case, only one roll with wavelength $\lambda_{TR}=4$ fills the domain, instead of two ``square'' rolls with $\lambda_{TR}=2$. This is consistent with the findings that the preferred $\lambda_{TR}$ increases with $Re$, and that large $Re$ simulations and experiments tend to find rectangular Taylor rolls with $\lambda_{TR}>2$, which however cannot be sustained at lower drivings\cite{ost14e} (cf. the experiments of Huisman \etal\cite{hui14} at $Re\sim\mathcal{O}(10^6)$ who found Taylor rolls with $\lambda_{TR}>3$).

\begin{figure}
  \centering
  \includegraphics[trim=0cm 1.5cm 0cm 2cm,clip=true,height=3cm]{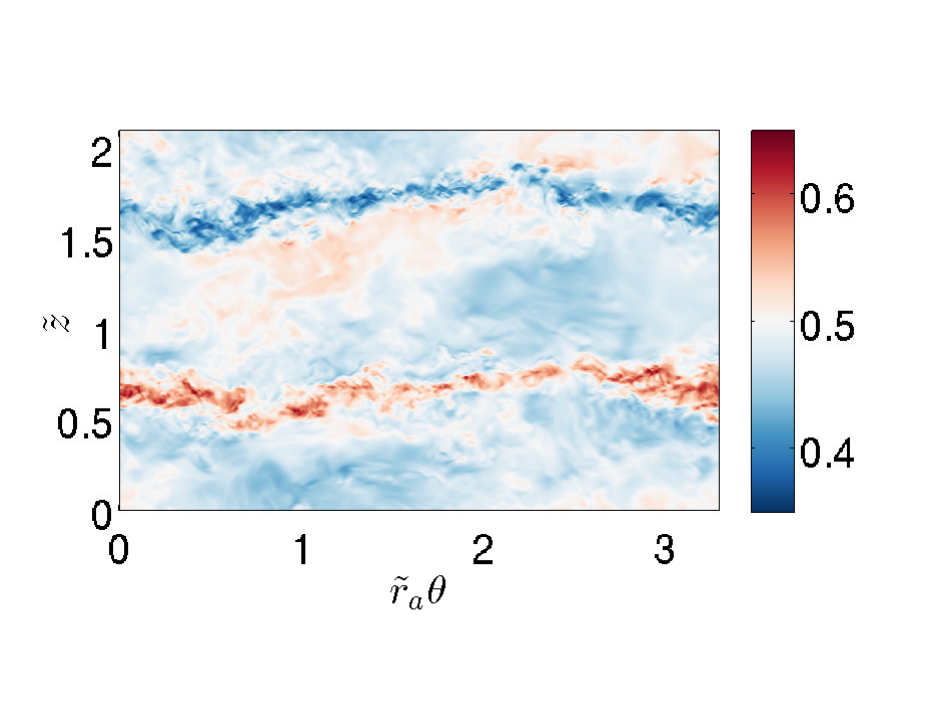}
  \includegraphics[trim=0cm 2.5cm 0cm 4cm,clip=true,height=3cm]{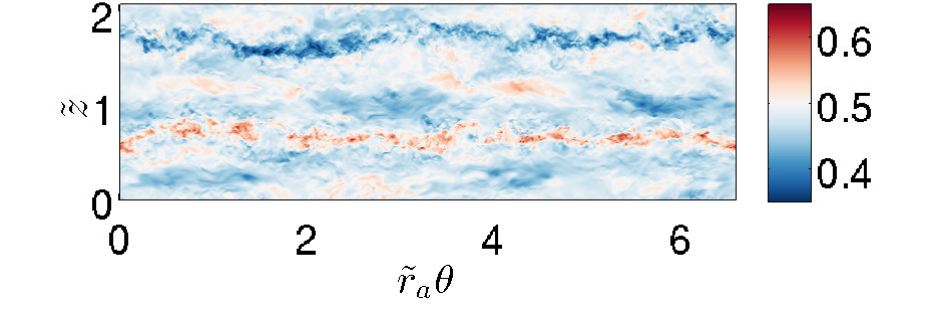}
  \caption{ Instantaneous azimuthal velocity $u_\theta$ at the mid-gap $\tilde{r}_a$ for the cases $\Gamma 2\text{N}20$ (left) and $\Gamma 2\text{N}10$ (right), as a function of the azimuthal variable $\theta$	. Increasing $n_{sym}$ allows larger wavelength structures to fit in the simulation. These structures azimuthally modulate the Taylor rolls, and allow for more and stronger fluctuations in the Taylor roll cores. }
\label{fig:q1insttheta}
\end{figure}

Long wavelength patterns, present in the axial direction, can also form in the azimuthal direction. Fig.~\ref{fig:q1insttheta} shows a pseudocolour plot of the instantaneous velocity $u_\theta$ at the mid-gap $\tilde{r}=0.5$ for the $\Gamma 2\text{N}20$ and $\Gamma 2\text{N}10$ cases. The axial signature of the Taylor rolls can be clearly appreciated in the panels. On top of this signature, additional azimuthal patterns can be seen. On the right panel of Fig.~\ref{fig:q1insttheta} structures similar to wavy Taylor vortices can be seen, which we will refer to as ``wavyness'' of the roll. These structures only appear in the $\Gamma 2\text{N}10$ case, but not in the $\Gamma 4\text{N}10$. This could be due to the increased distance between the plume clustering regions in the $\Gamma 4\text{N}10$, as the simulation domain is larger. We note that unlike Taylor rolls, long wavelength azimuthal structures are not stationary in time, as they are convected with the mean flow velocity and do not show up on temporal averages.

\begin{figure}
  \centering
  \includegraphics[width=0.49\textwidth]{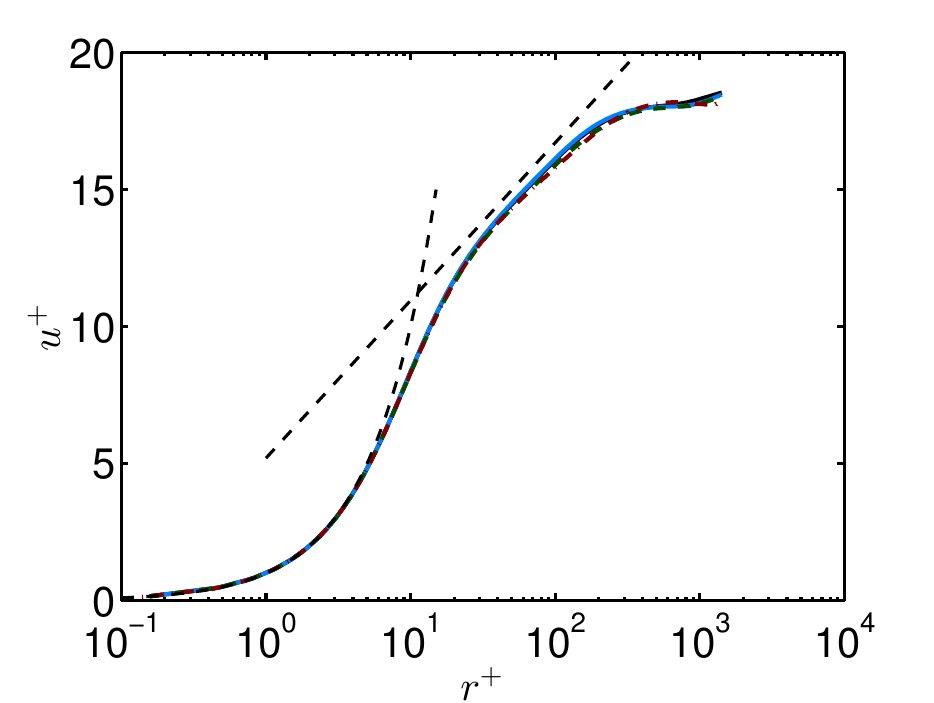}
  \includegraphics[width=0.49\textwidth]{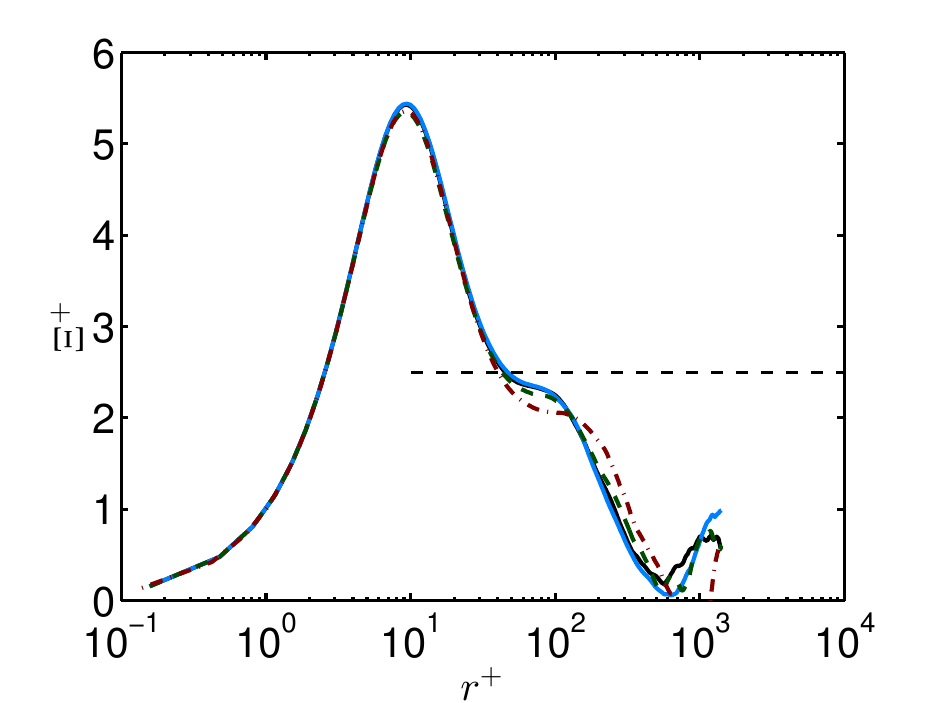}
  \caption{ The left panel shows the axially-, azimuthally- and time-averaged azimuthal velocity in inner cylinder wall units. The dashed lines indicate $u^+=r^+$ and $u^+=2.5\log r^+ + 5.2$. The right panel shows the diagnostic function $\Xi^+=du^+/d(\log r^+)$. Very little to no dependence on the box-size can be seen near the boundary layers, while differences can be appreciated from $r^+>500$, i.e. in the bulk. In the bulk, $u_\theta$, has a strong axial dependence, and this is probably causing the discrepancies. For both panels, symbols are as in Table~\ref{tbl:final}. }
\label{fig:q1wall}
\end{figure}

We will now quantify the effect of these patterns and thus of the computational box size on the flow. We start with one-point statistics, in particular with the mean azimuthal velocity. The left panel of Fig.~\ref{fig:q1wall} shows the azimuthally-, temporally- and axially- averaged azimuthal velocity $\langle u_\theta\rangle_z$ at the inner cylinder in inner cylinder wall units, while the right panel of Fig.~\ref{fig:q1wall} shows the so-called diagnostic function  $\Xi^+=du^+/d(\log r^+)$. which is the local slope of a lin-log plot of the profile. 

Only little variation between all cases can be seen between these panels. In the right panel, a slight decrease of the intercept of the logarithmic profile is observed for the $\Gamma 3\text{N}20$ and $\Gamma 4\text{N}10$ simulations. This is caused by the remnant axial dependence of $u_\theta$, which increases with increasing $\Gamma$. On the other hand, there are no appreciable differences in the boundary layers between the $\Gamma 2\text{N}20$ and $\Gamma 2\text{N}10$ simulations indicating that the azimuthal extent of the box is enough to capture the mean profiles in the boundary layer.

\begin{figure}
  \centering
  \includegraphics[width=0.49\textwidth]{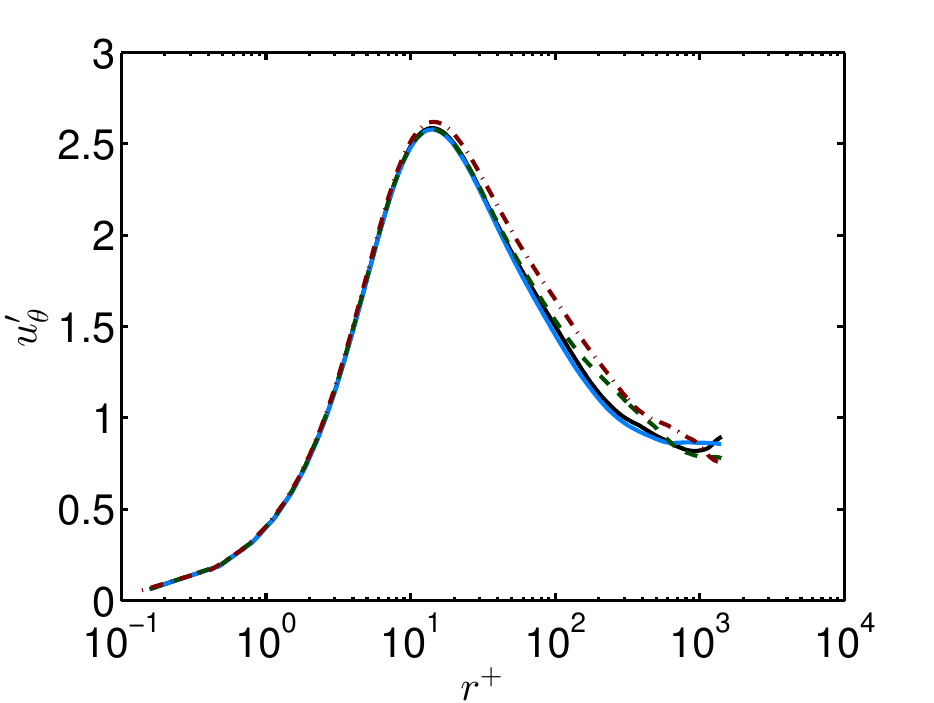}
  \includegraphics[width=0.49\textwidth]{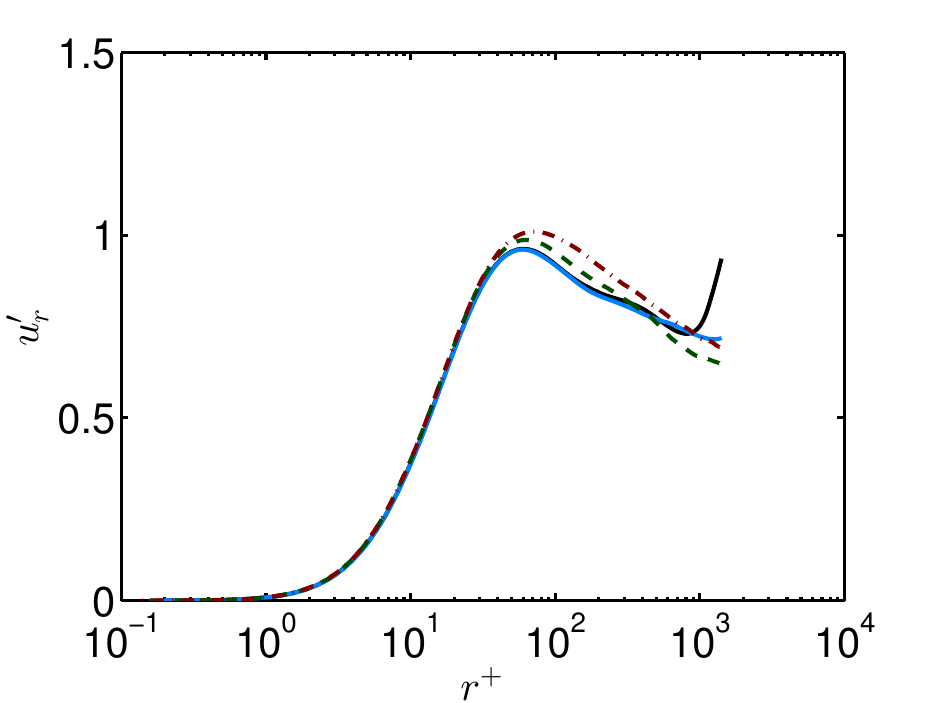}
  \includegraphics[width=0.49\textwidth]{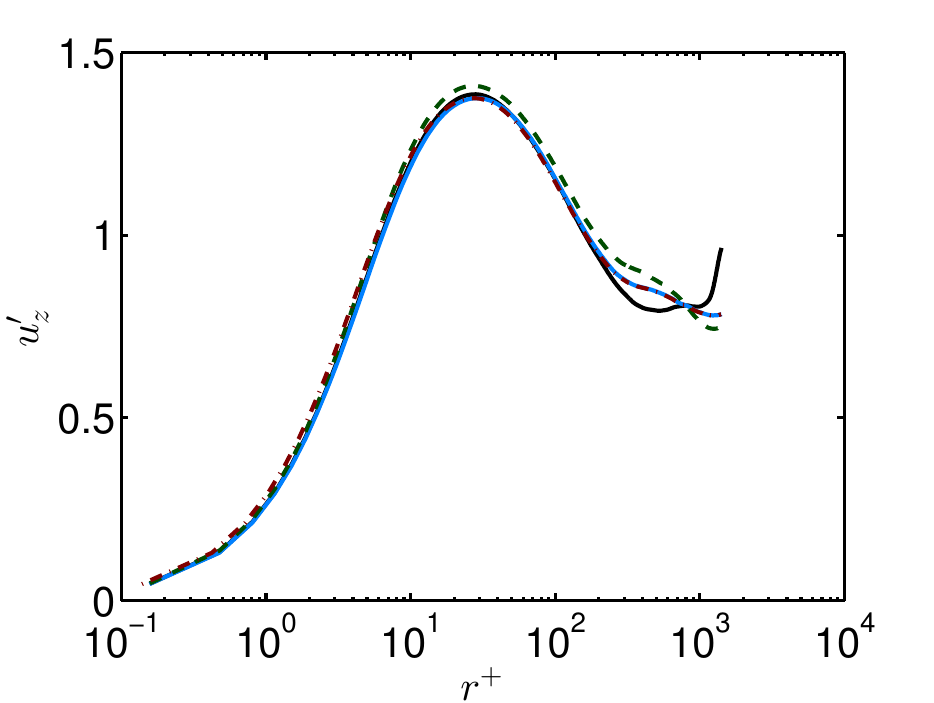}
  \includegraphics[width=0.49\textwidth]{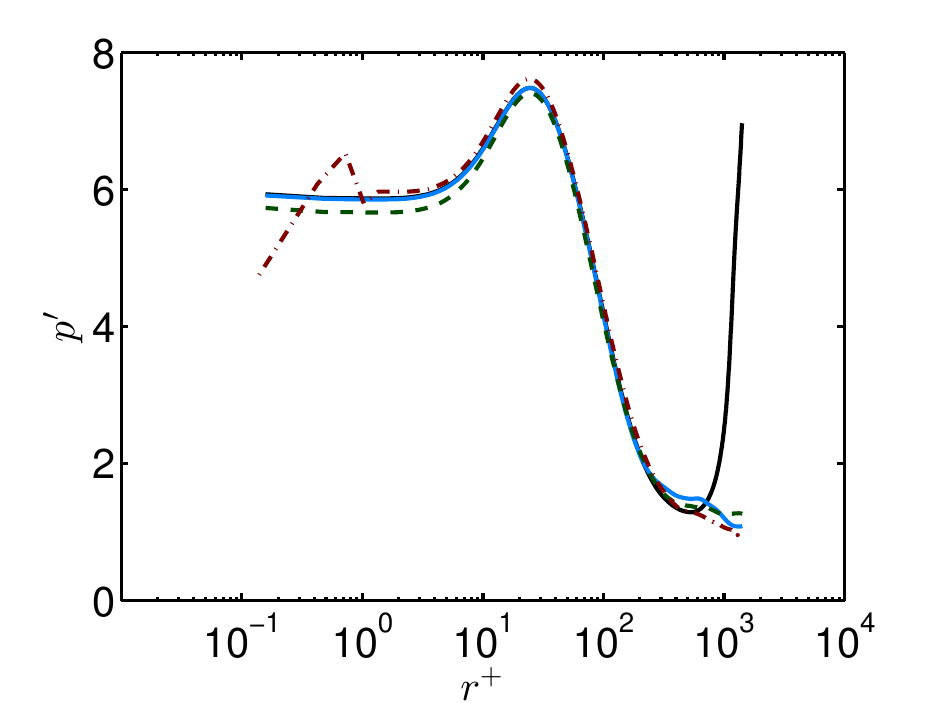}
  \caption{ Velocity and pressure fluctuations near the inner cylinder in inner cylinder wall units for all simulations.  Symbols are as in Table~\ref{tbl:final}. }
\label{fig:qrms}
\end{figure}

Fig.~\ref{fig:qrms} shows the velocity fluctuation profiles at the inner cylinder. The root mean square (rms) of a field $\phi$ is computed as $\phi^\prime=\langle \langle \phi^2\rangle_{\theta,t}-\langle\phi\rangle^2_{\theta,t}\rangle_z$. Note that the order of the axial average and subtraction operations is crucial, due to the remnant and significant axial dependence in the mean velocity fields, (cf. Fig.~\ref{fig:q1me}). Axially averaging before subtracting results in rms values which are considerably higher, but originate simply from the Taylor rolls and have nothing to do with the underlying statistics.

While the box-size can be seen to play a small effect on the $u_\theta^+$ profile, it is critical for other averages else. In general, increased box sizes lead to increased fluctuations, in line with what is seen in channels \cite{loz14}. The effect of the wavy patterns seen in Fig.\ref{fig:q1insttheta} on the fluctuations can be appreciated by comparing the $\Gamma2\text{N}10$ and the $\Gamma2\text{N}20$ cases. There is a clear increase in the $u_r^+$ and $u_z^+$ fluctuations in the bulk. This is a direct reflection of the increased mobility of the rolls. The axial extent of the domain can also be seen to affect the velocity fluctuation profiles. A larger axial domain, i.e. increasing $\Gamma$, again leads to larger fluctuations in the boundary layer, due to the increased mobility of the plumes, and the increased axial dependence of the velocity.

Finally, the pressure fluctuations become \emph{smaller} for larger domains. This is probably due to the pressure playing a damping role on the velocity fluctuations in the smaller domains. Again, this is in line with what is seen in channels \cite{loz14}. The geometric (computational) constraint on the flow inside the boundary layers is enforced through these pressure fluctuations. The sharp increase of velocity fluctuations in the bulk seen for the $\Gamma2\text{N}10$ case also results in an sharp increase of pressure fluctuations in the bulk.

\begin{figure}
  \centering
  \includegraphics[width=0.49\textwidth]{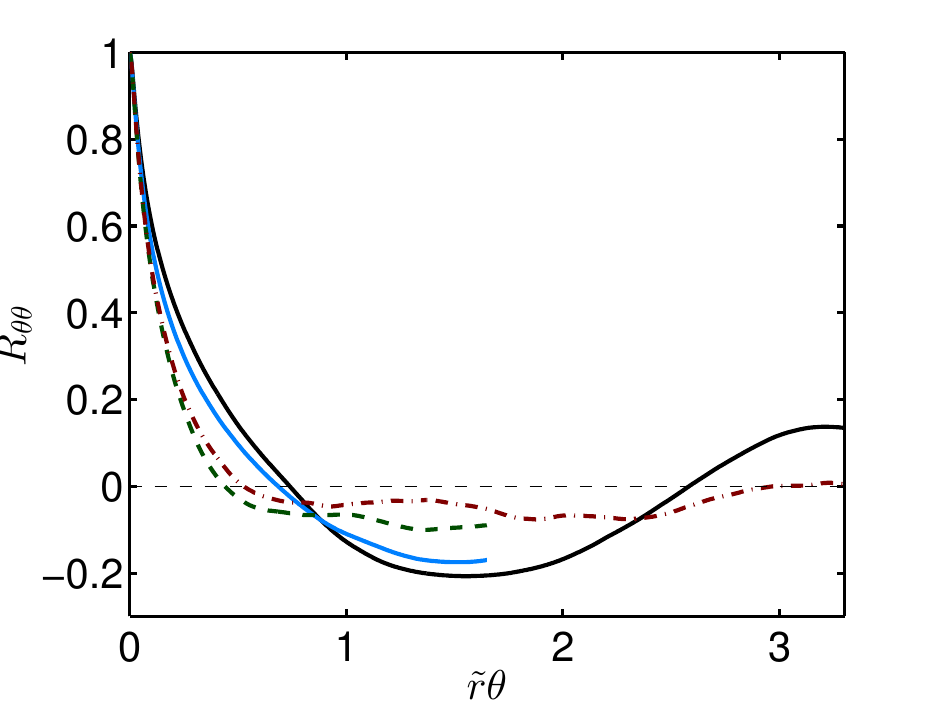}
  \includegraphics[width=0.49\textwidth]{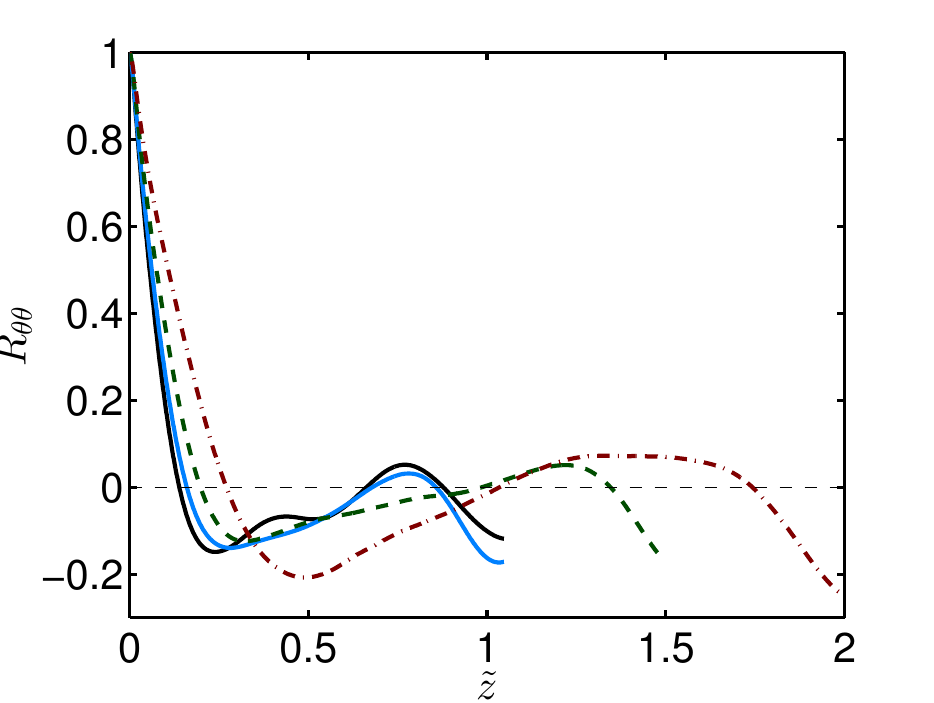}
  \includegraphics[width=0.49\textwidth]{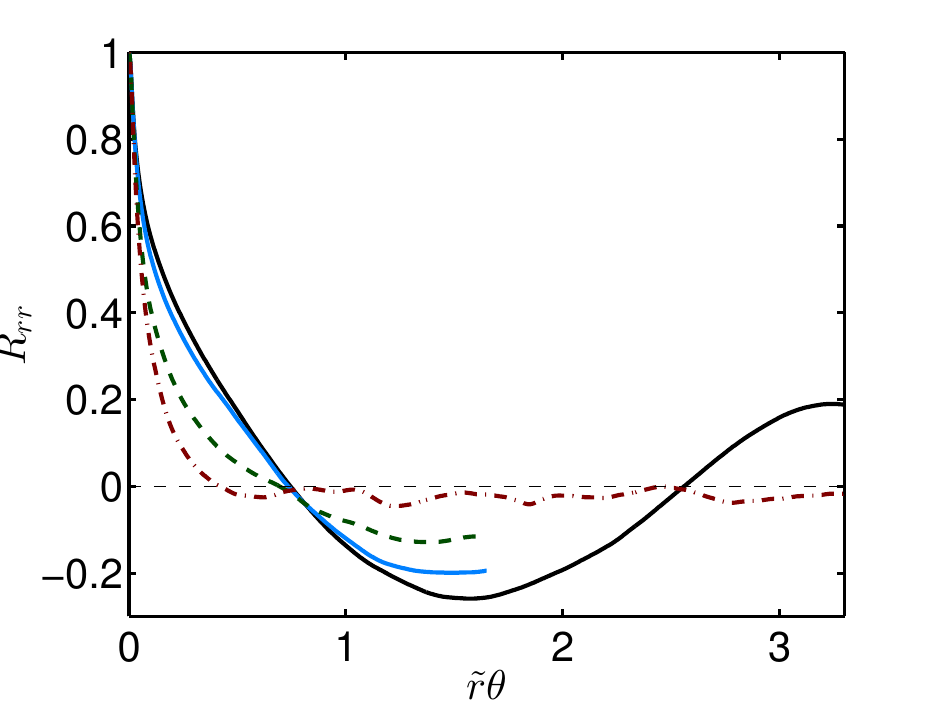}
  \includegraphics[width=0.49\textwidth]{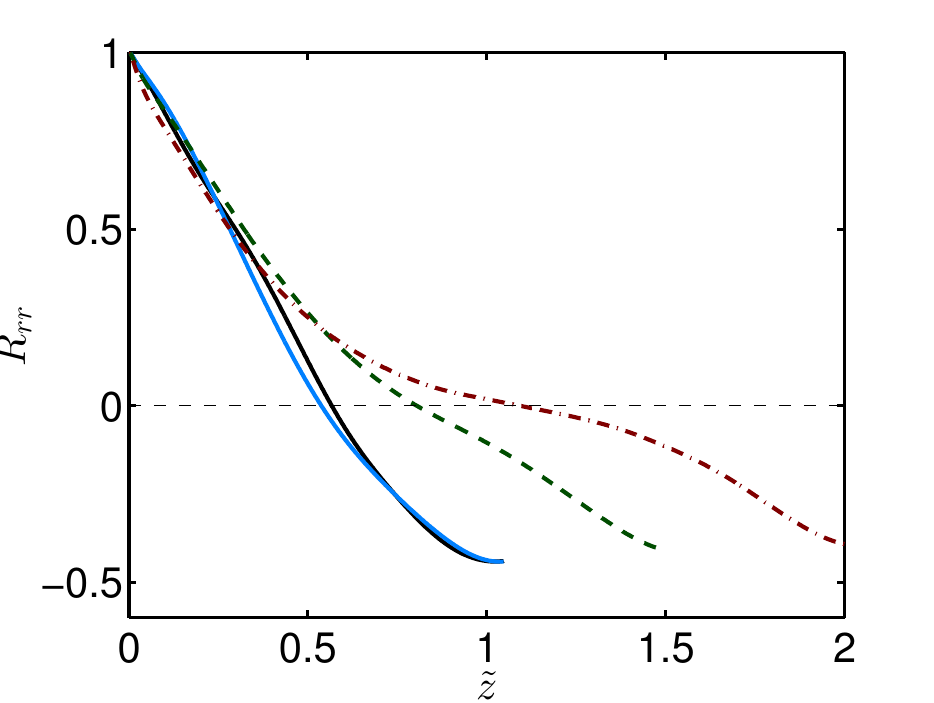}
  \includegraphics[width=0.49\textwidth]{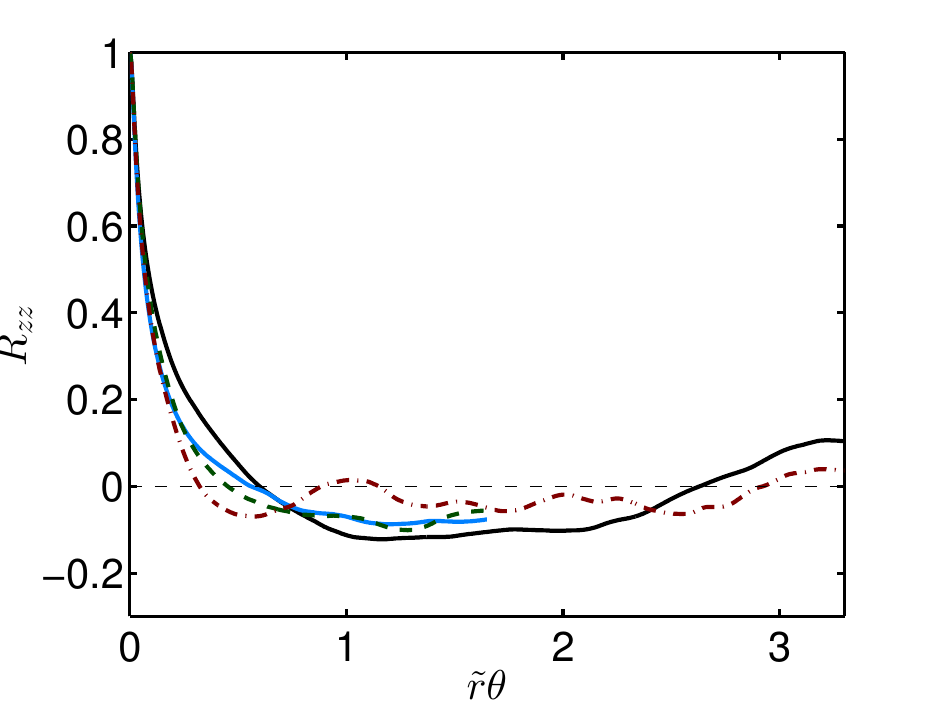}  
  \includegraphics[width=0.49\textwidth]{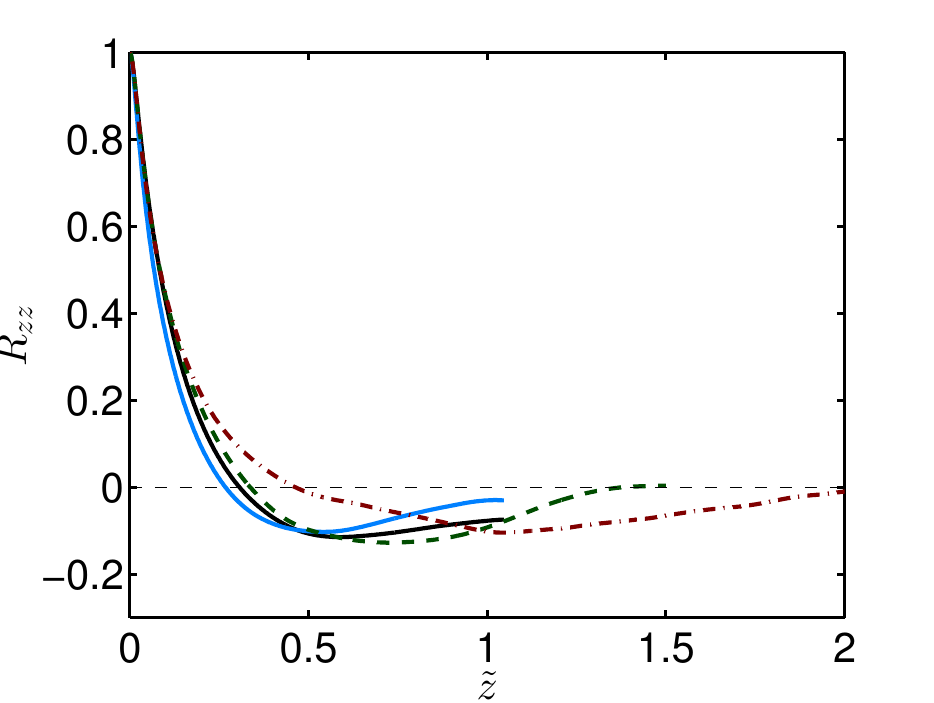}
  \caption{ Two-point autocorrelation functions at the mid-gap ($\tilde{r}=0.5$) for all simulations. Panels on the left column are autocorrelations in the azimuthal direction ($\theta$), while those on the right are autocorrelations in the axial directions. The three rows are for each velocity component: azimuthal (top), radial (middle) and axial (bottom). Symbols are as in Table~\ref{tbl:final}. }
\label{fig:autocorr}
\end{figure}

Fig.~\ref{fig:autocorr} shows the two-point autocorrelation function for all velocity fields in both axial and azimuthal direction. Two main effects of the box size can be seen on the flow. While the azimuthal extent of the box plays a negligible role in the decorrelation in all panels (compare $\Gamma2\text{N}10$ to $\Gamma2\text{N}20$ cases), the axial extent plays an important role in \emph{both} axial and azimuthal correlations. As expected, the axial autocorrelations are dominated by the effect of Taylor rolls. This is especially true in the case of the radial velocity autocorrelation $R_{rr}$. The axial velocity autocorrelation $R_{zz}$ remains relatively unaffected, as axial velocities in the mid-gap are very small (cf. Fig~\ref{fig:q1insttheta}). Additionally, increasing $\Gamma$ allows for a faster drop of $R_{\theta\theta}$ in the azimuthal direction, but not for faster drops in $R_{rr}$ and $R_{zz}$. This is due to the larger cores of the Taylor rolls, which result in more mixing. In these regions, the radial and axial velocities are small, so $R_{rr}$ and $R_{zz}$ are dominated by the strongly correlated regions seen in Fig.~\ref{fig:stst4cut}.

We also note that the azimuthal decorrelation lengths are an order of magnitude smaller than those seen in plane Couette flow\cite{bec95}, which have allowed TC simulations to reach higher $Re_\tau$ with heavily reduced computational costs. Current state-of-the art plane Couette simulations ``only'' reach $Re_\tau=550$ while requiring 2.2 billion points \cite{avs14}.

\begin{figure}
  \centering
  \includegraphics[width=0.49\textwidth]{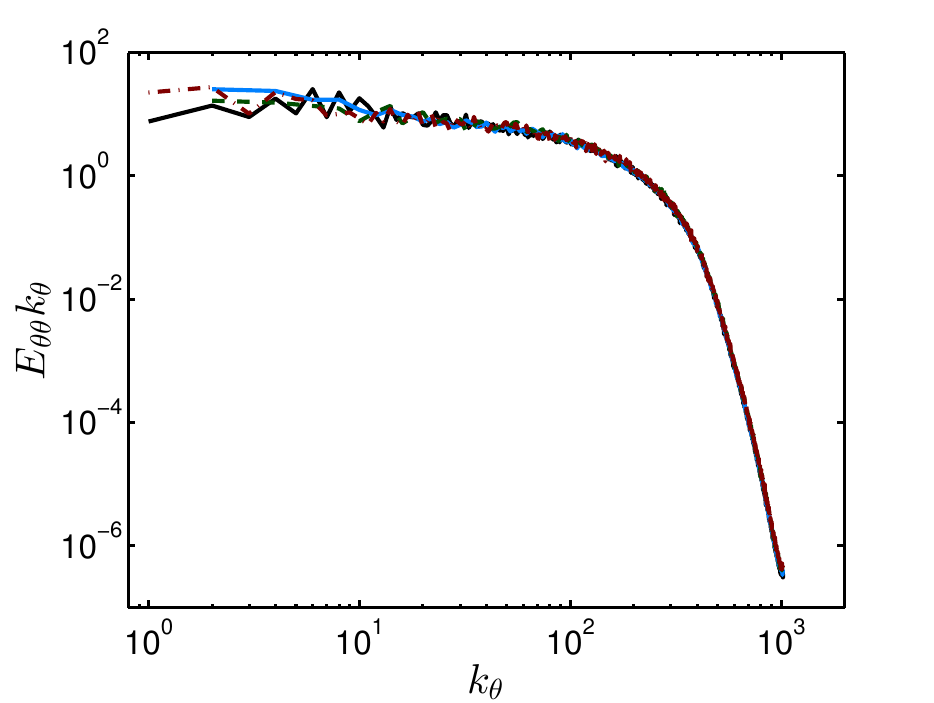}
  \includegraphics[width=0.49\textwidth]{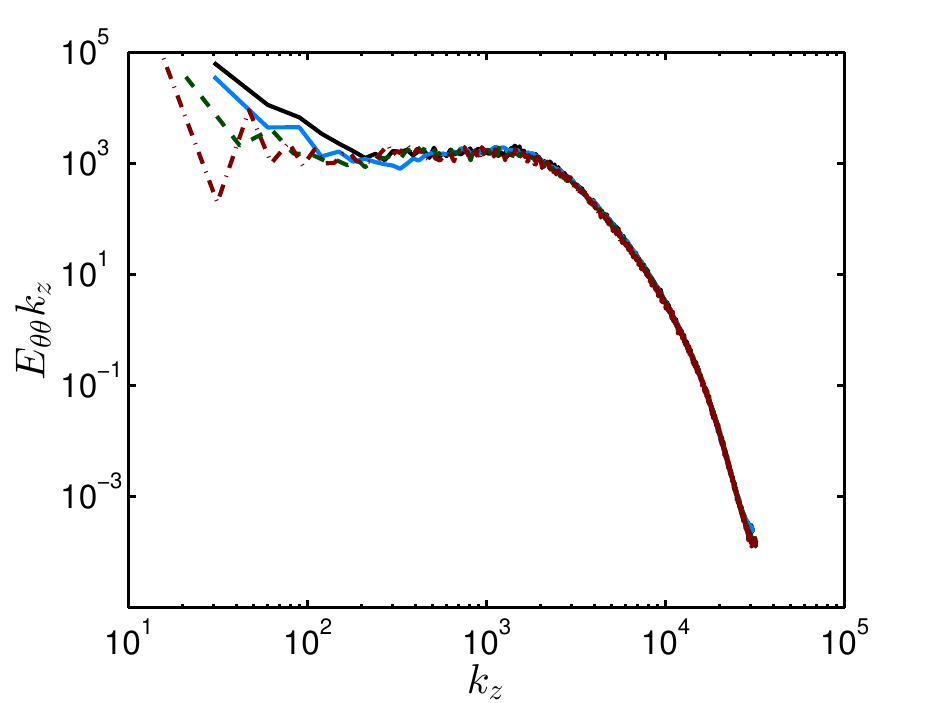}
  \includegraphics[width=0.49\textwidth]{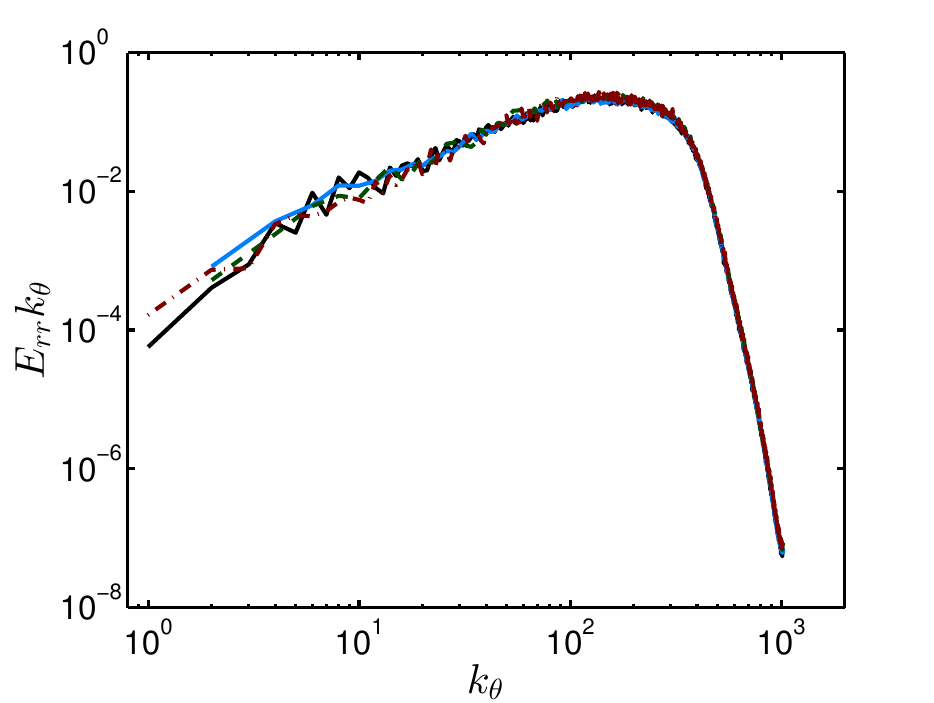}
  \includegraphics[width=0.49\textwidth]{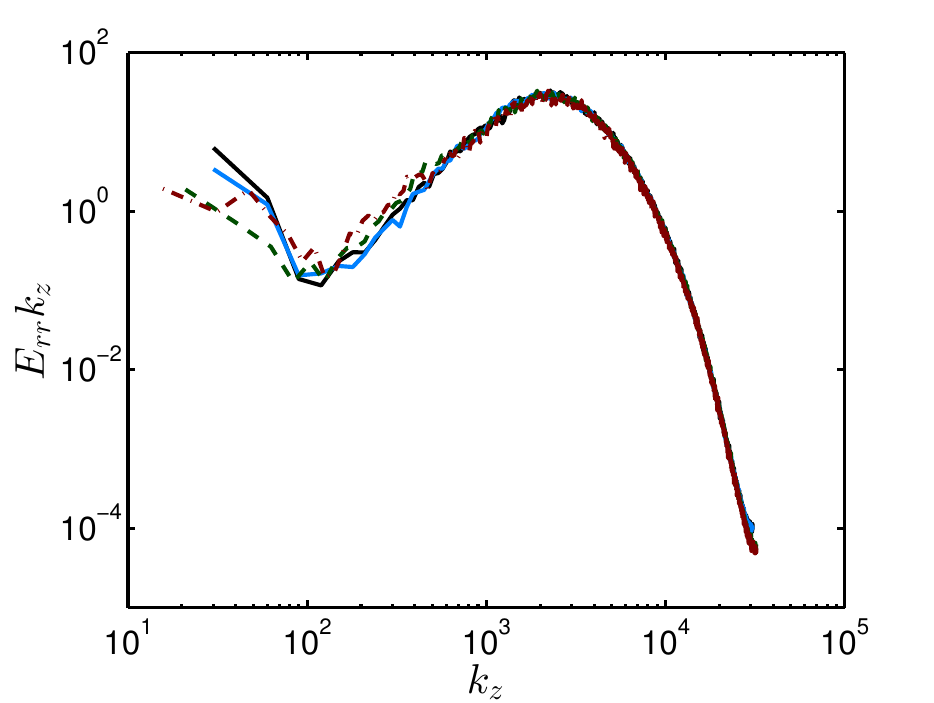}
  \includegraphics[width=0.49\textwidth]{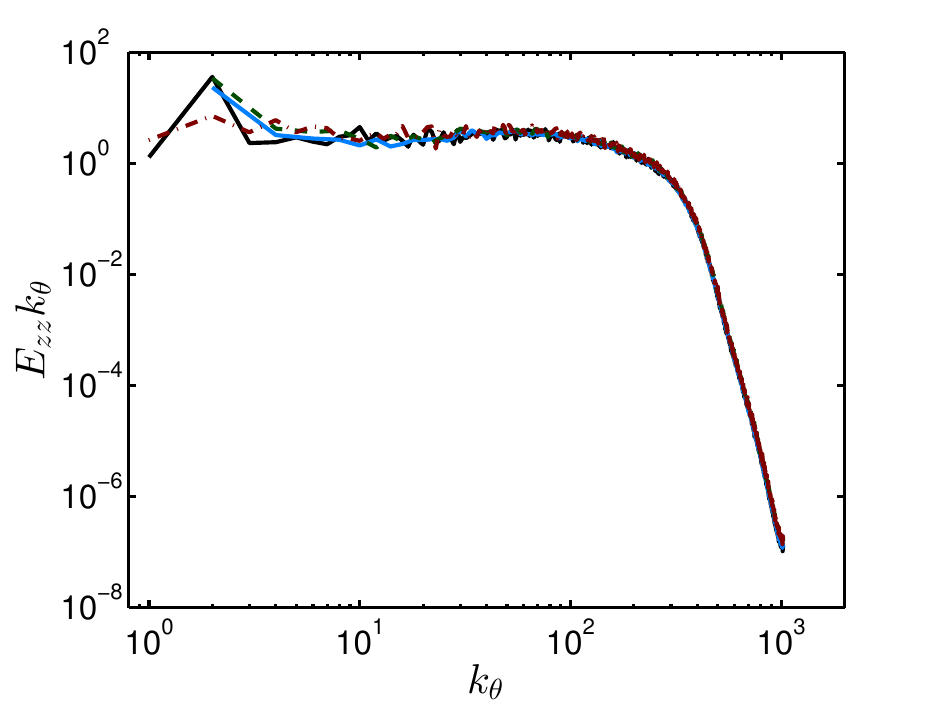}
  \includegraphics[width=0.49\textwidth]{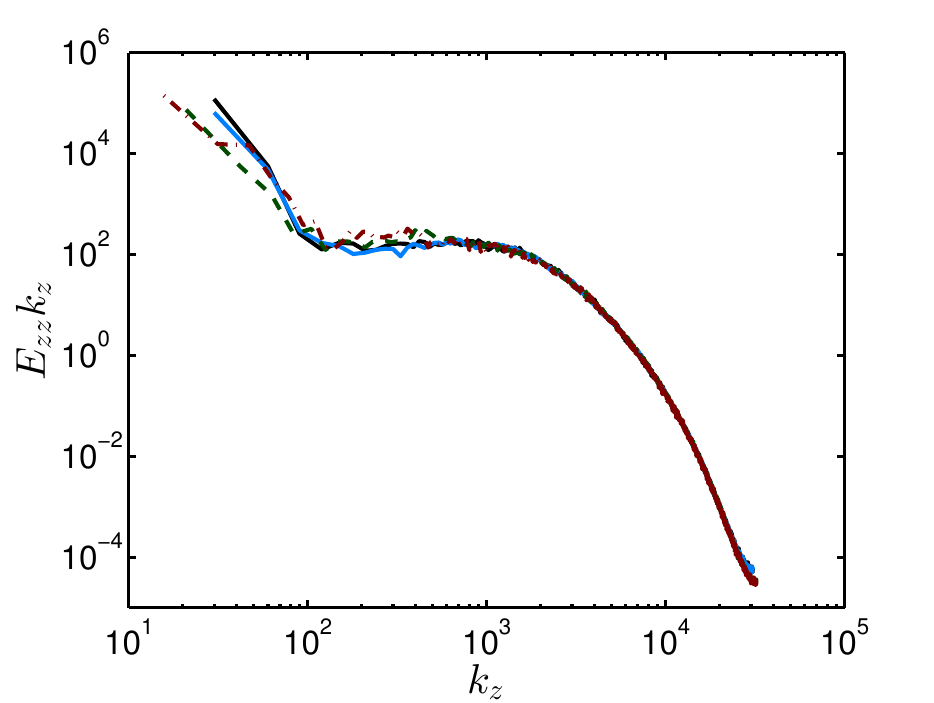}
  \caption{ Premultiplied azimuthal (left column) and axial (right column) spectra for the azimuthal (top), radial (middle) and axial (bottom) velocities at $r_i^+\approx12$. Symbols are as in Table~\ref{tbl:final}. }
\label{fig:spectraslab2}
\end{figure}

\begin{figure}
  \centering
  \includegraphics[width=0.49\textwidth]{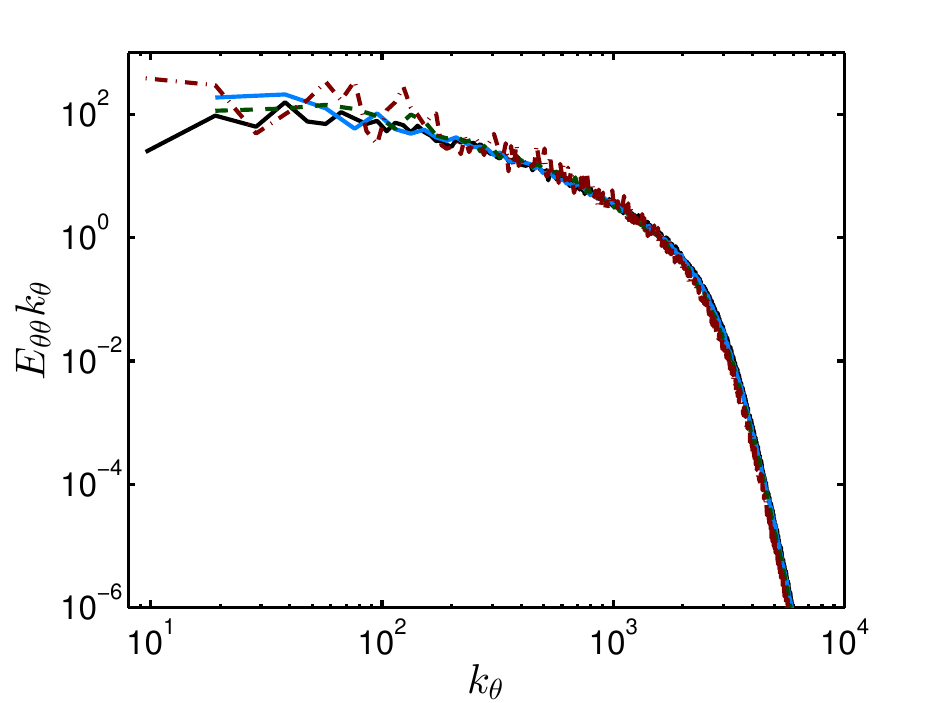}
  \includegraphics[width=0.49\textwidth]{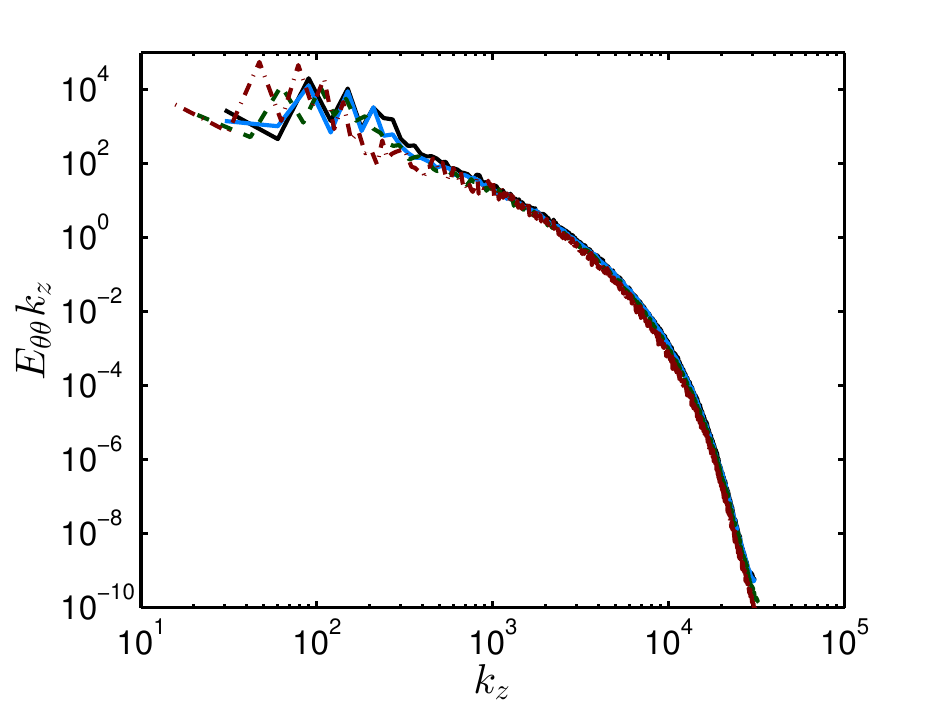}
  \includegraphics[width=0.49\textwidth]{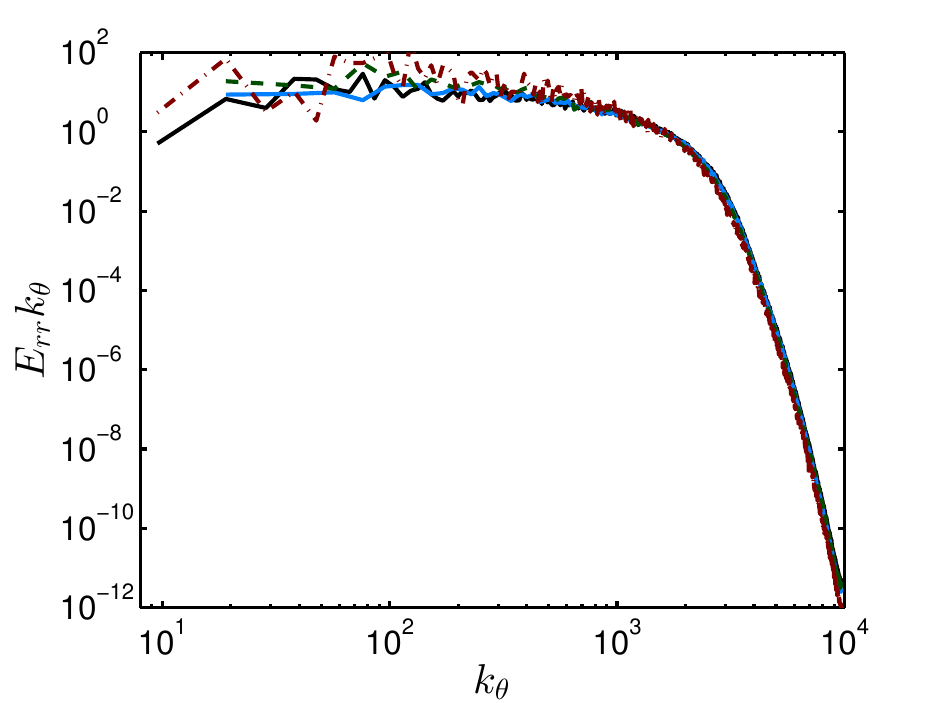}
  \includegraphics[width=0.49\textwidth]{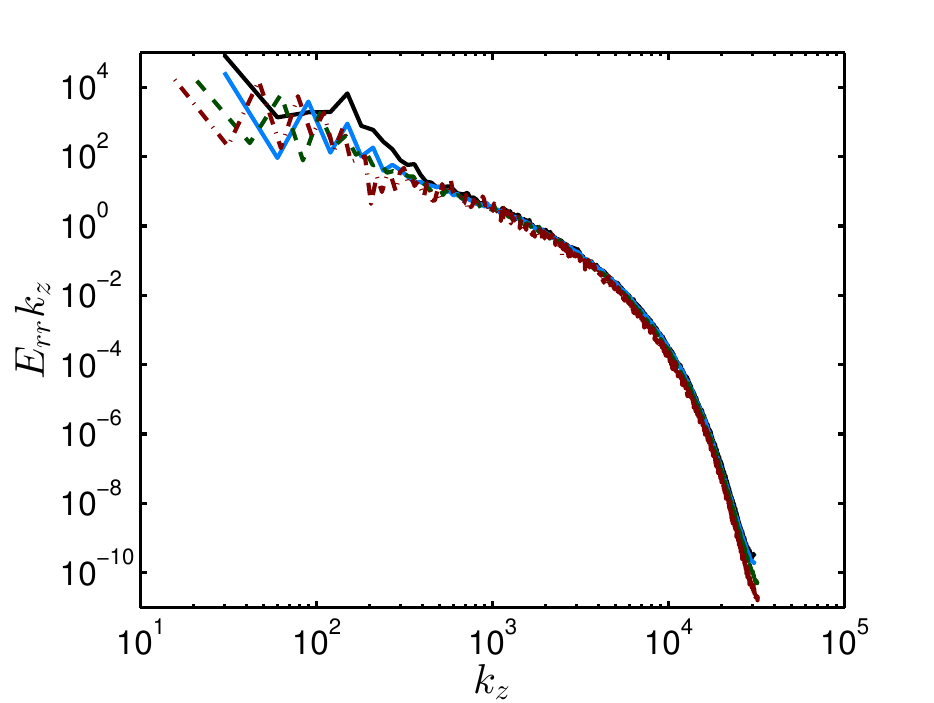}
  \includegraphics[width=0.49\textwidth]{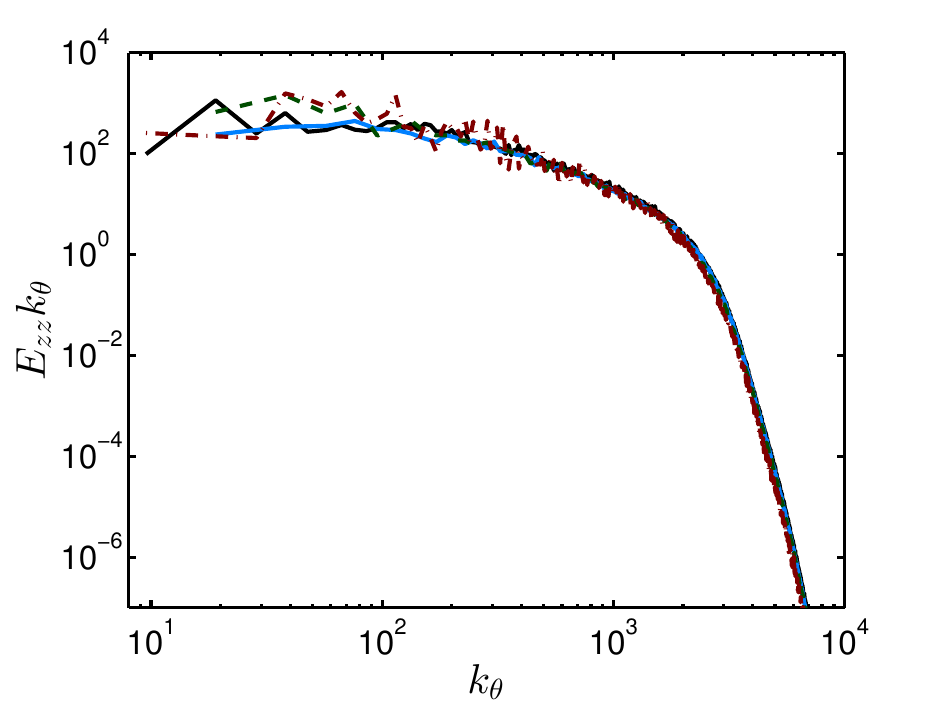}
  \includegraphics[width=0.49\textwidth]{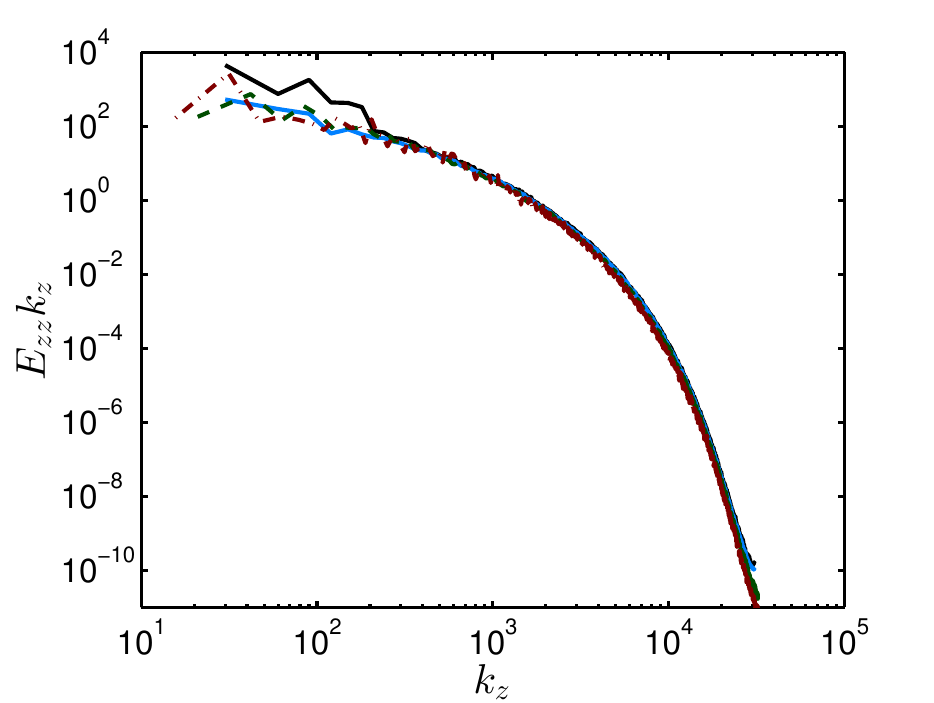}
  \caption{ Premultiplied azimuthal (left column) and axial (right column) spectra for the azimuthal (top), radial (middle) and axial (bottom) velocities at the mid-gap $\tilde{r}=0.5$. Symbols are as in Table~\ref{tbl:final}. No significant effect of the box size is seen for the small wavelength (large $k$) scales, while for the large scales, the azimuthal size of the domain appears to play a role in the axial structure of the rolls. The ``wavy'' modulations seen in Fig.~\ref{fig:q1insttheta} are clearly reflected in the spectra. }
\label{fig:spectraslab9}
\end{figure}

We now turn to the velocity spectra. Fig.~\ref{fig:spectraslab2} shows the premultiplied velocity spectra in the inner cylinder boundary layer ($r^+\approx12$), while Fig.~\ref{fig:spectraslab9} shows the spectra at the mid-gap ($\tilde{r}=0.5$). In both figures, the size of the computational box can be seen to play a negligible role for the spectra at the small scales, while, as seen in Ref.~\cite{ost15}, the large scales contain a very significant amount of energy both deep inside the boundary layer and in the bulk.

Inside the boundary layer there are two main energy containing scales- that of the plumes at high $k$, and that of the Taylor rolls, at low $k$. In the mid-gap, the plumes have merged with each other, and the main energy content can be found only in the large scales. It is apparent from both figures, that the maximum of the spectra is not converged to their very large box-size value. The box is not large enough to contain all large scales which are energetic. Even then, good collapse for the small scales can be seen. The largest scales contain energy from all three velocity components, and not only for the azimuthal and spanwise components, which is the case in channels\cite{hoy06} and plane Couette\cite{avs14}. All mid-gap spectra neither display the clear inertial range Kolmogorov scaling with $-5/3$ scaling, nor the $-1$ scaling for the $E_{\theta\theta}$ in the $\theta$-direction predicted by Perry \& Chong \cite{per86}. This is consistent with the experimental findings of Lewis \& Swinney \cite{lew99} and those of Huisman \etal\cite{hui13b}.

Saw-tooth patterns, indicating preferred even or odd modes can be seen for the axial spectra for all cases. Increasing $\Gamma$ shifts the maxima in $k_z$ accordingly, to accommodate for the different size of the Taylor-roll. We note that inside the boundary layer, simulations with larger $\Gamma$ have a smaller energy peak at the Taylor-roll wavelength $k_z$. This means that the Taylor-roll has a larger effect on the plume generation and its aspect ratio is smaller. Using a small domain may artificially strengthen the roll and lead to an increased correlation inside the boundary layer. Further increases in $\Gamma$ will accommodate for more rolls. This was seen to produce a very sharp dropoff in the spectra for $k_z<k_{TR}$ by Dong\cite{don07}.

Remarkably, sawtooth behaviour at low frequencies is also present in the azimuthal direction if the azimuthal extent of the domain is increased- indicating that by using a reduced azimuthal extent the formation of the wavy patterns is not allowed for. Further proof of this can be seen in Fig.~\ref{fig:spectraslab9}d) and \ref{fig:spectraslab9}f), where the energy of the large scales of the radial and axial velocity increases for the $\Gamma2\text{N}10$ case when compared to the $\Gamma2\text{N}20$  and $\Gamma3\text{N}20$ cases. Again, the spectra are not saturated. Extending the azimuthal extent of the simulation seems to be a necessity to fully capture the energy containing scales. This point requires further study, as due to the natural finiteness of the azimuthal extent, it could be  the case that the maximum in the spectra is at the longest wavelength if $n_{sym}=1$.

In summary, a systematic study of the effect of the computational box size on TC DNS was performed. From previous studies \cite{bra13}, it was already known that small boxes can obtain accurate results for the non-dimensional torque. Furthermore, similar to what was found by Lozano-Dur\'{a}n \etal \cite{loz14} for DNS of channel flow, small boxes also have accurate mean azimuthal (streamwise) velocity profiles in the boundary layers. Larger boxes are needed in order to obtain box-independent results for fluctuation values, two-point autocorrelations and low-wavelength spectra. The artificial truncation of the spectra by using a reduced box does not bring about significant changes in its structure at low wavelengths- even if the most energetic scales are not accounted for. Azimuthally small boxes do not allow for azimuthal wavy patterns in the Taylor rolls, and thus show a reduced level of fluctuations, as well as missing energy in the large scales for the axial and azimuthal spectra. 

In the axial direction, things are different. The size of the underlying Taylor roll dominates the autocorrelations, especially for the radial velocity. As we mentioned previously, the effect of a the computational box-size in the axial direction is not purely numerical, the wavelength of the Taylor vortex is a physical parameter. In experimental and natural realizations of TC, the wavelength of the Taylor rolls is determined by the axial constraints, i.e. end plates or periodicity. It appears from recent work \cite{mar14,ost14e} that for larger $\Gamma$ domains, which can accommodate more than a pair of rolls, the preferred wavelength of Taylor rolls increases with increasing $Re$. This effect cannot be captured in small box simulations, but can lead to bifurcations at high Reynolds numbers \cite{hui14}. Future work should consider simulating large $\Gamma\sim\mathcal{O}(10)$ to check the window of coexistence of different states at high $Re$, and how these affect the flow.

In the near future, a study of larger TC boxes seems mandatory, as to determine the minimal box size for accurate statistics of the velocity and pressure fluctuations and higher order moments. It can be the case that the axial extent is too small and thus non-physical, and that the azimuthal extent is not enough to develop wavyness. It is also unclear whether this wavyness is a product of artificial confinement in the axial direction, as they were only shown by the $\Gamma2\text{N}10$ and not by the $\Gamma4\text{N}10$. These increased fluctuations may provide a way for the system to overcome the energy barrier, and to switch between vortical states, i.e.\ from two vortex pairs with $\lambda_{TR}=2$ to one vortex pair with $\lambda_{TR}=4$. Another reason for doubting their physicality comes from the correlations, as the $\Gamma=2$ cases show unusually large decorrelation lengths in the azimuthal direction, and this might cause the formation of said patterns.

We also point out that the size of this minimal box is larger than those required for channels. The largest box in this manuscript, for which it is not clear yet whether its statistics are box-independent, has relative dimensions of $8$ half-gap lengths in the axial (spanwise) direction and $4.2\pi$ in the azimuthal (streamwise) direction, while accurate statistics were obtained for a box of size $\pi$ half-gaps in the spanwise direction and $2\pi$ half-gaps in the streamwise direction in Lozano-Dur\'{a}n \etal\cite{loz14}. 

\emph{Acknowledgments:} We would like to thank E. P. van der Poel for various stimulating discussions. RO would like to thank J. Jim\'{e}nez for motivating this study, and his group for the discussions during his stay in Madrid. We gratefully acknowledge an ERC Advanced Grant and the PRACE project 2013091966 resource CURIE based in France at Genci/CEA.

\bibliography{literatur}

\end{document}